\documentstyle[twocolumn,floats,psfig,aps]{revtex}
\def\Tau{{\cal T}}
\def\expval#1{{\langle #1 \rangle}}
\def\bra#1{{\langle #1 \vert}}
\def\ket#1{{\vert #1 \rangle}}
\def\etal{{\em et al}}
\def\chtr{\ket{\Psi_{ch.tr.}}}
\def\bchtr{\bra{\Psi_{ch.tr.}}}
\def\coh{\ket{\Psi_{coh.}}}
\def\cohphi#1{\ket{\Psi_{coh.}(#1)}}
\def\bcoh{\bra{\Psi_{coh.}}}
\begin{document}

\twocolumn[\hsize\textwidth%
\columnwidth\hsize\csname@twocolumnfalse\endcsname

\title{Electron Correlation and Charge Transfer Instability \\ in
Bilayered Two Dimensional Electron Gas}
\author{Sergio Conti$^{(1,2)}$ and Gaetano Senatore$^{(3,2)}$}
\address{$^{(1)}$Scuola Normale Superiore, Piazza dei Cavalieri,
I-56126 Pisa, Italy\\
$^{(2)}$Istituto Nazionale di Fisica della Materia\\
 $^{(3)}$Dipartimento di Fisica Teorica, Universit\`a di Trieste,
 Strada Costiera 11, I-34014 Trieste, Italy}
\date{September 27, 1996} 
\maketitle

\begin{abstract}
We prove that the predicted charge transfer state in symmetric
bilayers of two dimensional electron gases is always unstable at zero
bias voltage, due to interlayer correlation and/or tunneling.  This
is most easily seen by resorting to a pseudospin formalism and
considering coherent states obtained from the charge transfer state
through rotations of the pseudospins.  Evidently, the charge transfer
state is stabilized by a sufficiently strong gate voltage, as found in
recent experiments. We show that a simple model, in which the layers
are strictly two dimensional, is able to account quantitatively for
such experimental findings, when correlation is properly included.
\end{abstract}
\vskip1pc]
\narrowtext

Two dimensional electron gas (2DEG) systems confined in semiconductor
space charge layers (e.g., Si inversion layers, GaAs heterojunctions,
and quantum wells) have provided for the last twenty years an ideal
laboratory for studying various electron--electron interaction effects
under almost ideal 2D jellium conditions.  The zero temperature phase
diagram has attracted a lot of attention since Wigner pointed out that
at low density electrons would crystallize to minimize the potential
energy.  A stable spin polarized phase at intermediate density was
predicted by Hartree-Fock (HF) calculations, as a consequence of
competition between kinetic and exchange energy.  The most recent
Diffusion Monte Carlo (DMC) simulations by Rapisarda and Senatore (RS)
\cite{FR96} confirm this picture and show that the 2DEG is
paramagnetic up to $r_s=20$, then undergoes a ferromagnetic transition
and eventually crystallizes at $r_s=34$. As usual, the coupling
strength $r_s$ is defined in terms of the areal density by
$n_{2D}=1/\pi (r_s a_B)^2$, $a_B$ being the Bohr radius.

Bilayered systems, in addition to enhanced correlations effects with
respect to single layers, have an additional degree of freedom---the
interlayer distance $d$--- and exhibit a more complex phase diagram.
Paralleling a rich discussion on the high magnetic field Fractional
Quantum Hall regime\cite[and references therein]{Moon95}, some novel
interaction-driven phases have recently been proposed for the zero
field case\cite{FR96,RW91,SW91,DN93,AL94,DT94,FS97,MacDonald88}.
 
Based on the well-known fact that the compressibility of the low
density 2DEG is negative, Ruden and Wu (RW) \cite{RW91} argued that
exchange and correlation could overcome the kinetic and Hartree energy
costs stabilizing a charge transfer state (CTS), in which one layer
contains all the electrons and the other is empty.  Their HF
computation, restricted to the ideal zero-tunneling symmetric case,
predicts stability of the CTS for
\begin{equation}
r_s>{3\pi(\sqrt2+1)\over16} \left(1+2{d\over a_B}\right) \simeq 1.42
\left(1+2{d\over a_B}\right),
\end{equation} 
which is well into the range of the experimentally attainable electron
densities for reasonable layer separations $d$ (e.g., with GaAs
parameters, $r_s=4$ or $n_{2D}=2\cdot10^{10} {\rm cm}^{-2}$ for
$d<90$\AA, according to Ref. \onlinecite{RW91}).  In fact higher
values of the coupling ($r_s\approx 20$) have been 
achieved\cite{SS96} working with holes, rather than electrons.  Note
that in the bilayer we relate the density/coupling parameter $r_s$ to
the total areal density.

Evidence for a CTS in GaAs/AlAs double quantum well structures in the
presence of bias voltage was recently
reported\cite{Katayama94,Ying95,Katayama95}.  The experimental setup
allowed to measure independently the charge contained in the two
layers as a function of bias voltage, via Shubnikov-de Haas
oscillations.  For low voltages the transferred charge was found to
depend linearly on $V^e$, in quantitative agreement with the
predictions of a simple equivalent circuit model, but for high enough
voltages $V_{g}\simeq 1V$ the charge of one layer dropped abruptly
from around $20\%$ of the total charge to 0.

Here we shall demonstrate that, for two dimensional electron layers
and at zero bias voltage, the CTS is unstable at any density and layer
separation $d$.  A convenient model Hamiltonian for the coupled
electron layers is obtained introducing a pseudospin variable
$\tau_i$, which labels the planes with the convention that
$\tau_i^{(z)}=1/2$ ($-1/2$) if electron $i$ is in the upper (lower)
plane. For fixed interlayer distance and background densities and in
the absence of tunneling, the Hamiltonian  takes the form
\begin{eqnarray}
H &=& \sum_i {p_i^2\over 2m} + {1\over2}{\sum_{i\ne j}}^{\prime} \left( {e^2\over
r_{ij}}{1+4\tau_i^{(z)}\tau_j^{(z)}\over2} + \nonumber\right.\\
&&\left.+{e^2\over
\sqrt{d^2+r_{ij}}}{1-4\tau_i^{(z)}\tau_j^{(z)}\over2} \right)
+V^e \sum_i \tau_i^{(z)},
\end{eqnarray}
where $V^e$ is the energy difference between the two planes, given by
$2\pi e^2 (\rho_1-\rho_2)d$ in the absence of external fields,
$\rho_1$ and $\rho_2$ being the background densities of the two
planes. The prime on the second sum indicates that the term with $k=0$
should be omitted when the sum is rewritten in reciprocal space.  It
is clear that with N electrons there are two CTS's, corresponding to
two eigenstates of the total pseudospin component along $z$,
$\Tau^{(z)}=\sum_i\tau_i^{(z)}=\pm N/2$. In other words CTS is the same
as (pseudospin) ferromagnetism, with the magnetization oriented along
the $z$ axis.  We shall show that without bias voltage there can be
no such ferromagnetic phase transition. If any, in the presence of
tunneling a transition takes place to a (coherent) correlated state,
which has the same symmetry as the charge transfer state but does not
displace any charge. The charge transfer phase transitions can take
place in the presence of significant sample asymmetry and/or gate
voltage.

It is interesting to consider the ideal case $d=V^e=0$, where both
real spin $\sigma$ and pseudospin $\tau$ are conserved and the system
is actually a 4-component monolayer 2DEG. At large density (small
$r_s$) the ground state is completely paramagnetic, with the four
components equally populated to minimize the Fermi energy; in the
opposite low density regime the ground state is known to be a Wigner
crystal and essentially insensitive to spin polarization.  In the
intermediate regime the 4-component 2DEG can be expected to mimic the
above-mentioned behavior of the 2DEG (and 3DEG), with the appearance
of a spin-polarized ground state. In HF, the 4-component 2DEG is
unstable towards the 2-component phase at $r_s=3\pi/16(\sqrt2-1)\simeq
1.42$ and then towards the 1-component phase at
$r_s=3\pi/8\sqrt2(\sqrt2-1)\simeq2.01$, the first being obviously
identical with the $d\to0$ limit of the RW results and the latter with
the usual spin polarization transition of the 2DEG in HF.

\begin{figure}
\caption{Ground state energy as a function of $r_s$ for the normal
unpolarized ($E_2$), polarized ($E_1$) and 4-component ($E_4$) 2DEG
and for the triangular Wigner crystal ($E_{sol}$). Except for $E_4$,
all data are from RS. On the purpose of clarity we plotted
$r_s^{3/2}(E(r_s)-c_1/r_s)$, with $c_1=-2.2122$.}
\label{fig1}
\psfig{figure=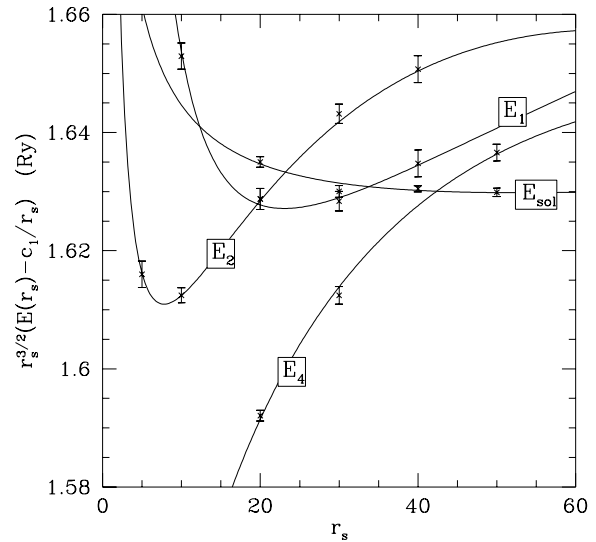,width=0.9\columnwidth}
\end{figure}

To assess the actual presence of these phase transitions we performed
Slater-Jastrow Variational and Fixed Node Diffusion Monte Carlo
simulations for the 4-component 2DEG, and compared them with the RS
results.  The method is completely analogous to the one employed by
RS, with the only difference that our DMC code moves all particles at
every step, therefore requiring smaller time steps which, in turn,
makes time step extrapolations unnecessary. To check the code we
duplicated one of the RS data points, finding excellent agreement
within statistical noise.  Our results are reported in Table
\ref{tabellamc}, and compared in Figure \ref{fig1} with the data by
RS on the 2 and 1-component 2DEG. At variance with the simple HF
prediction, the figure clearly shows that the 4-component phase
crystallizes at $r_s\simeq 42$, being always stable at smaller
couplings with respect to the other two fluid phases that we have
considered.  One could argue that backflow corrections might spoil the
validity of our results, but we do not think this to be the case since
backflow corrections are known to lower the paramagnetic energies more
than the ferromagnetic ones.

As we have already mentioned, a total charge transfer state is
described in our formalism as an eigenstate of each $\tau_i^{(z)}$
with eigenvalue $+1/2$. We remark that $\tau_i^{(z)}$ commutes with
the Hamiltonian even at finite $d$ and $V^e$ (but zero tunneling) and
we label with $\chtr$ the lowest eigenstate within the charge transfer
subspace.  In the symmetric $V^e=0$ case it can be easily seen that
the coherent state $\coh=\exp\left\{i {\pi\over2}\sum_j
\tau_j^{(y)}\right\}\chtr$ has an average energy lower than $\chtr$.
In fact, (i) $1/x > 1/\sqrt{d^2+x^2}$ for any $x$; and (ii) $\bchtr
4\tau_i^{(z)}\tau_j^{(z)}\chtr=1$, whereas
$\bcoh4\tau_i^{(z)}\tau_j^{(z)}\coh=0$.  The variational principle
implies that $\chtr$ cannot be the ground state of the system, even if
$\coh$ is not an eigenstate of $H$.

We stress that both $\chtr$ and $\coh$ have the same symmetry
properties under exchange of the particle's spatial coordinates, as
this is not affected by pseudospin rotation. In particular the two
states are both fully pseudospin polarized, though in different
directions. The average charge distribution in $\coh$ corresponds to
half electrons in one layer and half in the other, and therefore it
does not loose in Hartree energy with respect to the normal state,
still having the exchange energy gain deriving from the full
antisymmetry of the spatial part of the wavefunction.  The difference
with the above-mentioned HF predictions is due to the fact that no
state with the same symmetry as $\coh$ was considered by RW.

No such clearcut statement can be made in the asymmetric $V^e\ne0$
case. In general, it can be seen that the term $V^e \sum_i
\tau_i^{(z)}$ favours finite values of $\Tau^{(z)}$, i.e. (partial)
pseudospin polarization. A quantitative 
estimate would rely on the detailed pseudospin dependence of the
correlation energy for the bilayer system, which --- to the best of our
knowledge --- has not yet been determined.  Common wisdom saying that
the phase transition in jellium models occurs directly from
unpolarized to fully polarized implies the presence of a maximum in
the internal energy at intermediate polarizations. In turn, this
implies that if the value of $V^e$ is sufficient to raise the
polarization up to this point, which may still be far below
saturation, it will spontaneously saturate. This behavior would look
like a 2-component to 1-component phase transition driven by an
increasing $V^e$, and can possibly explain the experimental results
reported in Refs.~\onlinecite{Katayama94,Ying95}.

It must be noticed that the state $\coh$ is a broken symmetry state,
since all states of the form $\cohphi{\phi}=\exp\left\{i
{\pi\over2}\sum_j
(\tau_j^{(y)}\cos\phi+\tau_j^{(x)}\sin\phi\right\}\chtr$ are
degenerate. The order parameter is the total pseudospin $\Tau=\sum_i
\tau_i$, corresponding to interlayer phase coherence, and behaves as
an easy-plane ferromagnet, the $z$ component being frozen by the
interplay of $V^e$ and the Hartree energy.  Most of the considerations
presented in Ref.~\cite{Moon95} for the high $B$ situation can be
directly extended to this zero field case.  The role of a tunneling
Hamiltonian, which can be conveniently written as
\begin{equation}
H_t=t \sum_i \tau_i^{(x)},
\end{equation} with  the tunneling
matrix element $t>0$ , is to break the $\phi$ symmetry and stabilize
the coherent state $\coh=\cohphi{\phi=0}$, in which
$\expval{\Tau^{(z)}}=0$ and $\Tau^{(x)}=-N/2$.  In the large
$t$ case all electrons trivially lie in the symmetric state, which has
exactly the same form as $\coh$.  For small enough layer separation
$d$ and tunneling amplitude $t$ a perturbative estimate of the
critical amplitude $t^*$ at which the fully coherent state $\coh$
becomes stable is easily obtained from the energies of Table
\ref{tabellamc} and of RS, on the ground that the $d$ dependence is negligible,
as it vanishes in first order, contrary to the tunneling term. Thus
one gets $t^*/2 = E_{4}(r_s)-E_{2}(r_s)$, which at $r_s=2$ corresponds
to $t^*= 0.8\, {\rm meV}=9 K$ with GaAs parameters.

The experimental setup of
Ref. \onlinecite{Katayama94,Ying95,Katayama95} includes a metallic
gate at a large distance $D$ from the two 2DEG's and some charged
dopants, whose location and amount are unspecified and irrelevant for
the present purposes.  The 2DEG's are realized with 150(180)\AA\ thick
GaAs quantum wells with a midpoint separation $d$ of 220(194)\AA\ 
for sample A(B), and the tunneling energy $t=0.005 K\,(5 K)$.  From the
work of Zheng and MacDonald\cite{ZM} and from numerical
simulations\cite{FS97} it is clear that interlayer correlations are
relevant only if $d$ becomes smaller than the average interparticle
distance $r_s a_B$. As in the present situation $r_sa_B \sim 150$\AA\
and $d\sim 200$\AA\, correlation effects can be safely neglected. In
the following we shall further neglect finite thickness and tunneling
effects.

We propose a simple model which correctly includes intralayer
correlations and interlayer mean field interactions and is closely
related to the one used by Eisenstein \etal\cite{Eisenstein94} to
discuss a similar experimental setup.  Let us consider the gate
potential $V_g^{(0)}$ corresponding to equal density $n^{(0)}$ in the
two `active' layers, i.e. to zero electric field between the
layers. [The value of $V_g^{(0)}$ depends on details of the sample
design, as location and amount of dopants, geometry, etc.].  With
reference to this state and under the assumption that no net charge
flows in the system with varying $V_g$, the electrochemical
equilibrium condition between the two 2DEG's is written as
\begin{equation}
\mu(n_1)-\mu(n^{(0)})=\mu(n_2)-\mu(n^{(0)}) + 4\pi e^2 d
(n_2-n_0)\end{equation}
and between the upper layer and the gate as
\begin{equation}
-e(V_g -V_g^{(0)})=
\mu(n_1)-\mu(n^{(0)}) +4\pi e^2 D (n_1+n_2-2n^{(0)}) 
\end{equation}
We now proceed to solve numerically these two coupled equations for
the unknowns $n_1$ and $n_2$ at various values of $V_g$.  While the
sample parameters $d$, $n^{(0)}$ and $V_g^{(0)}$ are given in
Ref.~\onlinecite{Ying95}, $D$ is not, and was determined by fitting
the measured variation of the average density $(n_1+n_2)/2$ with
$V_g$.

\begin{figure}
\caption{Layer densities for the bilayer A specified in the text, as
a function of the bias voltage. Triangles and diamonds give the
experimental results of Ref.~\protect\cite{Ying95}. The full curve
results from the strictly two dimensional model discussed in the text,
while the dashed curve is obtained neglecting both correlation and
exchange.  The inset reports the pseudospin polarization $\xi$ as a
function of bias potential $V_g$.}
\label{fig2}
\psfig{figure=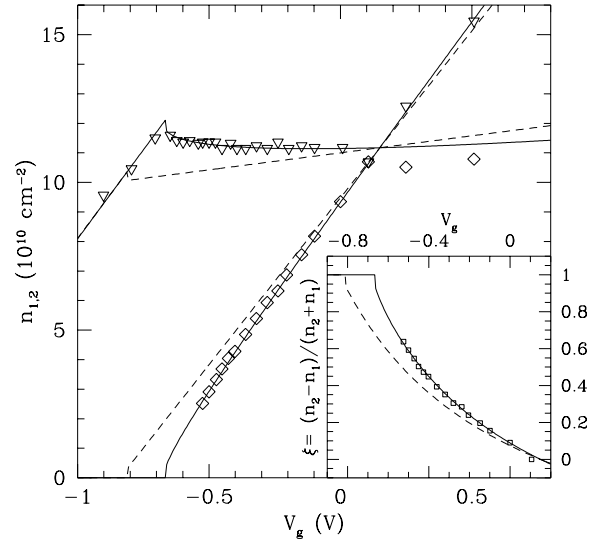,width=0.9\columnwidth}
\end{figure}

\begin{figure}
\caption{Layer densities for the bilayer B specified in the text, as
a function of the bias voltage. Triangles and diamonds give the
experimental results of Ref.~\protect\cite{Ying95}. The full curve
results from the strictly two dimensional model discussed in the text,
while the dashed curve is obtained by accounting for tunneling as
explained in the text.}
\label{fig3}
\psfig{figure=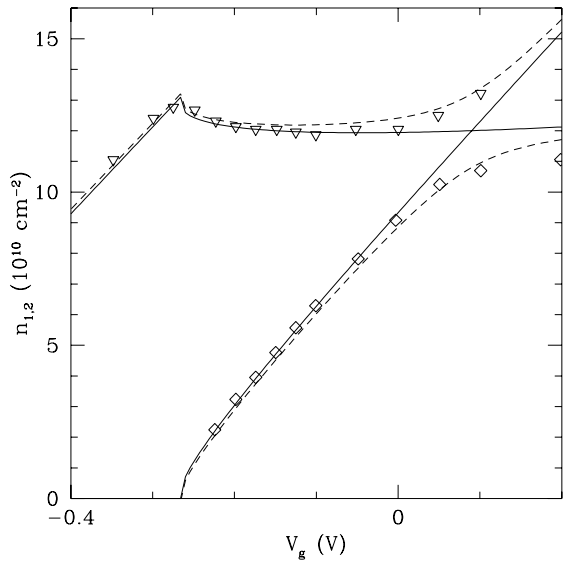,width=0.9\columnwidth}
\end{figure}

The chemical potential $\mu(n)=d(nE(n))/dn$ for the idealized single
layer 2DEG is derived from the precise equation of state obtained by
RS with DMC.  The results shown in Figs. \ref{fig2}, \ref{fig3} are in
very good agreement with the experimental data from
Ref.~\onlinecite{Ying95}, with significant disagreement only in the
tunneling-dominated region around $n_1=n_2$. In fact quantitative
agreement (see Fig. \ref{fig3}) can be obtained even in this region
computing the subband densities from the layer densities
with no tunneling $\tilde n_1$, $\tilde n_2$ by exact
diagonalization of a noninteracting electron hamiltonian with finite
tunneling,
\begin{eqnarray}
H&=&\sum_k \left[ \left(\frac{\hbar^2k^2}{2m} -
\mu^0(\tilde n_1)\right)a^{\dag}_{k,1}a^{}_{k,1} +\nonumber\right.\\
&&\hskip-.8cm\left.\left(\frac{\hbar^2k^2}{2m} - 
\mu^0(\tilde n_2)\right)a^{\dag}_{k,2}a^{}_{k,2}
+t\left(a^{\dag}_{k,1}a^{}_{k,2} 
+c.c.\right)\right].
\end{eqnarray}
We remark that neglecting intralayer interactions would lead to the
usual approximation $\mu^0(n)=C_qn$, where the so-called quantum
capacitance $C_q$ is given by $\pi \hbar^2/m$. By making this
assumption for both layers the above equations are linear and can be
solved analytically, obtaining results in fair agreement with
experiments but completely missing the nontrivial charge transfer
effect signaled by the peak in the upper curve of Fig.\ref{fig2}.

In conclusion, we presented a pseudospin formalism to describe bilayer
2DEG systems and used it to rigorously prove the instability of the
charge transfer state and to argue for the stabilization of coherent
states at small finite tunneling.  We also reported results from DMC
simulations, which reveal the stability of the pseudospin unpolarized
state at $d=V^e=0$. Finally we have shown that the presently available
experimental data, being in a regime where interlayer correlations are
negligible, can be quantitatively accounted for by a simple strictly
two dimensional model.  Clearly the more challenging and interesting
situations, in which interlayer correlations are discernible and
determine the details of the charge transfer, is still awaiting for
both experimental and theoretical investigations.

A preliminary account of the above study was presented at the {\it
INFM-FORUM workshop on 2DEG}, held in Pisa, FORUM, June 96, and at
the Conference {\it The Electron Quantum Liquid in Systems of Reduced
Dimensions} held in Trieste, ICTP, July 96. We acknowledge stimulating
discussions with B. I. Halperin and with A. H. MacDonald.

{\em Note added}---While completing the present manuscript we became
aware of a recent preprint on the same topic by Das
Sarma \etal\cite{DasSarma}, which is somewhat complementary to this
work.

\newpage\mediumtext 
\begin{table}
\caption{Fixed-node DMC total energies of the 4-component
2DEG for 52 particles and in the bulk limit, in Ryberg per particle.
Variational results used for the size-extrapolation are also
given. 
These data are accurately reproduced by the same fitting formula used
by RS, with parameters $a_0=-0.88115$, $a_1=4.439$, $a_2=0.14063$, 
$a_3=1.9034$.}
\label{tabellamc}
\begin{tabular}{cccccc}
& $r_s=2$ & $r_s=10$ & $r_s=20$ & $r_s=30$ & $r_s=50$ \\ \hline
$N=36$    & -0.5882(4) & -0.17102(2) & -0.092275(6) & -0.063561(5) & -0.039386(3) \\ 
$N=52$    & -0.5728(1)& -0.17030(2) & -0.092085(4) & -0.063474(3) & -0.039354(1)\\ 
$N=84$    & -0.58062(7)& -0.17055(2) & -0.092123(5) & -0.063481(4) & -0.039355(2)\\
$N=100$   & -0.5758(1)& -0.17036(2) & -0.092062(4) & -0.063457(4) & -0.039344(2)\\
$N=180$   & -0.57676(8)& -0.17031(2) & -0.092050(5) & -0.063453(4) & -0.039340(2)\\
$VMC_{\infty}$
	  & -0.5750(1) & -0.17019(3) & -0.092002(4)& -0.063426(3) & -0.039330(1)\\
\hline	
$DMC_{52}$& -0.5838(4) & -0.17232(4) & -0.092891(9) & -0.063975(8) &-0.039639(3)\\
$DMC_{\infty}$
          & -0.5860(5)& -0.17221(6) & -0.09281(1) & -0.063927(9) & -0.039615(4)\\
\end{tabular}
\end{table}
\end{document}